\PassOptionsToPackage{breaklinks,colorlinks,citecolor=blue}{hyperref}
\documentclass[twocolumn]{aastex631}

\shorttitle{Eccentric triples}
\shortauthors{Mangipudi, Grishin et al.}

\usepackage{graphicx}	
\usepackage{amsmath}	
\usepackage{amssymb}	
\usepackage{color}
\usepackage{float}

\begin{document}

\title{Extreme eccentricities of triple systems: Analytic results}

\correspondingauthor{Evgeni Grishin}
\email{evgeni.grishin@monash.edu}

\author{Abhi Mangipudi}
\affil{School of Physics and Astronomy, Monash University, VIC 3800, Australia}

\author[0000-0001-7113-723X]{Evgeni Grishin}
\affil{School of Physics and Astronomy, Monash University, VIC 3800, Australia}
\affil{OzGrav: Australian Research Council Centre of Excellence for Gravitational Wave Discovery, Clayton, VIC 3800, Australia}

\author[0000-0001-5371-3432]{Alessandro A. Trani}
\affil{Department of Earth Science and Astronomy, College of Arts and Sciences, The University of Tokyo, 3-8-1 Komaba, Meguro-ku, Tokyo 153-8902, Japan}
\affil{Okinawa Institute of Science and Technology, 1919-1 Tancha, Onna-son, Okinawa 904-0495, Japan}
\affil{Research Center for the Early Universe, Graduate School of Science,
The University of Tokyo, 7-3-1 Hongo, Bunkyo-ku, Tokyo 113-0033, Japan}

\author[0000-0002-6134-8946]{Ilya Mandel}
\affil{School of Physics and Astronomy, Monash University, VIC 3800, Australia}
\affil{OzGrav: Australian Research Council Centre of Excellence for Gravitational Wave Discovery, Clayton, VIC 3800, Australia}

\begin{abstract}

Triple stars and compact objects are ubiquitously observed in nature. Their long-term evolution is complex; in particular, the von-Zeipel-Lidov-Kozai (ZLK) mechanism can potentially lead to highly eccentric encounters of the inner binary. Such encounters can lead to a plethora of interacting binary phenomena, as well as stellar and compact-object mergers. Here we find explicit analytical formulae for the maximal eccentricity, $e_{\rm max}$, of the inner binary undergoing ZLK oscillations, where both the test particle limit (parametrised by the inner-to-outer angular momentum ratio $\eta$) and the double-averaging approximation (parametrised by the period ratio, $\epsilon_{\rm SA}$) are relaxed, for circular outer orbits. We recover known results in both limiting cases (either $\eta$ or $\epsilon_{\rm SA} \to 0$) and verify the validity of our model using numerical simulations. We test our results with two accurate numerical N-body codes, \textsc{rebound} for Newtonian dynamics and \textsc{tsunami} for general-relativistic (GR) dynamics, and find excellent correspondence.  We discuss the implications of our results for stellar triples and both stellar and supermassive triple black hole mergers.  
\end{abstract} 

\keywords{Close binary stars (254) -- Multiple stars (1081) -- Supermassive black holes (1663) -- Stellar dynamics (1596) -- Stellar mass black holes (1611) -- Galaxy mergers (608) -- Compact objects (288)
}


\section{Introduction}\label{intro}

Triple and multiple systems are ubiquitous for diverse scales and architectures: from Solar-system asteroids \citep{margot2015} and Kuiper belt multiples \citep{noll,grishin2010, roz2020}, to planets around binary stars and multiple planetary systems \citep{winn2015}, to stellar triples and multiples \citep{Tokovinin:2006,raghavan2010, duchene2013}. 

Although the general three-body is notoriously chaotic \citep{poincare, ValtonenBook2006}, in hierarchical systems, where the inner binary is perturbed by a distant companion, perturbative techniques allow certain triple systems to be integrable. In particular, the von Zeipel-Lidov-Kozai (ZLK) mechanism \citep{vZ1910,lid62,koz62} can generate highly eccentric binaries that lead to close encounters and mergers. ZLK had been invoked almost ubiquitously as a central agent in the evolution of triple and multiple systems on planetary \citep[e.g.][]{wu2003,FT07, pn09, naoz11, naoz12, hamers16, hamers17}, stellar \citep[e.g.][]{mazeh79, Eggleton2001, per_fab, naoz14, toonen2020, toonen21, hamers21, hamers22, grishin22}, galactic \citep[e.g.][]{ant12, hoang18, Fragione2019} and extra-galactic \citep[e.g.][]{bonetti16, man21} scales.  For example, the recent discovery of a massive, compact and stable stellar triple TIC 470710327 \citep{tess_triple} serves as a unique laboratory for triple dynamics \citep{vg22}. ZLK may also provide an important channel for driving compact object binaries toward merger as gravitational-wave sources (see, e.g., \citealt{MandelFarmer:2018,Mapelli:2021,MandelBroekgaarden:2021} for recent reviews).  In this context, ZLK can play a key role in isolated field triples or quadruples 
\citep[e.g.,][]{silsbee2017, antonini2017, HamersThompson:2019, fragione2020, trani2022}.  It may also impact the evolution of merging double compact objects in in dense environments such as globular clusters \citep{MillerHamilton:2002b,Antonini:2016} or galactic nuclei \citep[e.g.,][]{hoang18, Fragione2019}. 

The ``standard" ZLK integrable mechanism relies on the double-averaging (DA) procedure, where the Hamiltonian is averaged over the mean anomalies of both binaries, such that the evolution occurs over long, \textit{secular}, timescales \citep{ford2000}, much longer than both orbital periods. 
In addition, when the interaction potential is truncated at the leading, quadrupole order, recent studies found analytic solutions in the test particle limit, when one of the inner binary members has low mass with respect to the other member \citep{kinoshita07, lubow21}, and in the general case \citep{hamers_kozai}.

When either the DA or the quadrupole approximation break down, the problem becomes chaotic again. The next leading order, the octupole order, applies when the outer orbit is eccentric (i.e. the eccentric ZLK mechanism, \citealp{katz2011,lithwixk2011}) and when the inner binary masses are unequal. The timescales of octupole effects are generally longer than the quadrupole timescale. 

The DA approximation can also break down and cause chaotic evolution, but it was given much less attention. Historical studies of the Lunar motion resolved the tension of additional apsidal precession in terms of additional ``evection terms" \citep{Tisserand}. The perturbation theory was generalised for irregular satellites on eccentric and inclined orbits \citep{cuk2004}. Recently, \citet{Luo2016} found an effective potential which corrects the DA procedure and makes it more compatible with N-body simulations, while still retains the attractive features of fast integration of the secular equations. \cite{grish17} used the additional corrections to generalise the Hill-stability limit for arbitrary inclination, and \cite{Grishin2018} found corrected formulae for the maximal eccentricity, $e_{\rm max}$ and the critical inclination in the mildly hierarchical case for the test particle limit.

Although the \cite{Grishin2018} extension is limited to cases where the tertiary is much more massive (the Hill approximation) on a circular orbit, direct N-body integrations of more realistic binaries produce more merger rates per galaxy and more eccentric mergers than secular integrations \citep{Fragione2019}.

Motivated by the enhanced merger rates of mildly-hierarchial systems, we take a further step in the analytic description in the mildly-hierarchical three body problem. We relax the test-particle approximation and find a formula for $e_{\rm max}$ for any dynamically stable triple system in an outer circular orbit. Our  expression of $e_{\rm max}$ reduces to known results in the literature for the relevant limits and allows analytical estimates for encounters and mergers of triple systems, with direct implications to triple stellar evolution and field triple BH mergers.
 
The paper is organised as follows. In Section~\ref{review} we review the preceding work on secular dynamics of triple systems. In Section~\ref{formalism} we motivate the need for and derive a new analytic formula for the maximal eccentricity and discuss its validity and limitations. In Section~\ref{results} we present our results, comparing the analytic expressions to three-body numerical simulations. We discuss the caveats ans astrophysical implications in Section~\ref{discussion}.  We summarise in Section~\ref{sum}. 

\section{Overview of secular dynamics}\label{review}

Here we review the governing equations of the standard ZLK mechanism. We then extend the overview to include non-secular contributions that affect the DA formalism as well as relativistic corrections.

\subsection{Standard von Zeipel-Lidov-Kozai mechanism}

Consider an inner binary with masses $m_0$ and $m_1$, semi-major axis $a$ and eccentricity $e$. The total mass is $m_{\textrm{bin}} = m_1 + m_0$. The inner binary is perturbed by an outer companion of mass $m_{\rm out}$, semi-major axis $a_{\rm out}$ and eccentricity $e_{\rm out}$. The normal to the outer orbital plane is denoted by $\hat{\textbf{n}}_{\rm out}$.  The inner binary's eccentricity vector is  $\boldsymbol{e}$ and its specific angular momentum vector (expressed in units of the circular angular momentum of a binary with the same orbital semi-major axis) is $\boldsymbol{j} = \sqrt{1 - e^2}\hat{\boldsymbol{j}}$.  The DA quadrupole term in the Hamiltonian (we will sometimes refer to this specific energy as the ``potential'') that governs the evolution of $\boldsymbol{e}$ and $\boldsymbol{j}$ is 

\begin{align} \label{phi-quad}
\Phi_{\textrm{Quad}} &= \frac{\Phi_0}{8}[1 - 6e^2 - 3(\boldsymbol{j}\cdot\hat{\boldsymbol{n}}_{\rm out})^2 + 15(\boldsymbol{e}\cdot \hat{\boldsymbol{n}}_{\rm out})^2], \nonumber \\
\end{align}

where 

\begin{align}
\Phi_0 &= \frac{Gm_{\textrm{out}} m_0m_1 a^2 }{m_{\textrm{bin}}a_{\textrm{out}}^3(1 - e_{\textrm{out}}^2)^{3/2}}.
\end{align}

We omitted terms that contain the semi-major axes, corresponding to the energies of the Keplerian ellipses, which are assumed to be constant. In terms of the argument of pericentre of the inner binary $\omega$ and the mutual inclination between the orbital planes $i$, using $j_z \equiv {\textbf{j}\cdot\hat{\textbf{n}}_{\rm out} = \sqrt{1 - e^2}\cos i}$ and $e_z \equiv {\textbf{e}}\cdot\hat{\textbf{n}}_{\rm out} = e\sin\omega\sin i$, the quadrupole term can be written as \citep{Naoz2016}

\begin{align}
\Phi_{\textrm{Quad}} &= \frac{\Phi_0}{8}[1 - 6e^2 - 3j_z^2 + 15e_z^2], \nonumber \\ & =\frac{\Phi_0}{8}[2 + 3e^2 - 3(1 - e^2 + 5e^2\sin^2\omega)\sin^2i_{\textrm{tot}}].
\end{align}

For large initial mutual inclinations, $|\cos i_0|<\sqrt{3/5}$, the inner eccentricity is coherently excited until a certain threshold (see Eq.~(\ref{maxDA}) in section \ref{formalism}).

Additional extensions to the standard ZLK mechanism are i) including higher-order terms in the multipole expansion, ii) breaking down the DA hierarchical approximation, iii) going beyond the test particle approximation, iv) including additional external forces. We neglect i) for two reasons: the timescales on which the higher-order terms contribute are longer and we focus on circular outer binaries, $e_{\rm out}=0$, which sets the octupole term to zero. Under this assumption, combining effects (ii),(iii) and (iv) allows us to find and verify a unified formula for the maximal eccentricity when these effects are taken into account together. We review each extension below.

\subsection{Corrected Double Averaging}

The DA approximation neglects perturbations on timescales shorter than the secular timescale \citep{kinoshita2007, Antognini2015}

\begin{equation}
\tau_{\rm sec} \approx \frac{1}{2\pi}\frac{m_{\rm tot}}{m_{\rm out}} \frac{P^2_{\rm out}}{P_{\rm in}} (1 - e^2_{\rm out})^{3/2},
\end{equation}
where $m_{\rm tot} = m_{\rm out} + m_{\rm bin}$.
When the system is only mildly hierarchical, the accumulated errors in neglecting these perturbations may be large. It is possible  to correct for these errors through the use of the single-averaged (SA) equations of motion \citep{Luo2016}, which depend on the position of the outer body on its orbit. The parameter that quantifies the level of the hierarchy and these short-timescale perturbations is the SA strength

\begin{equation}
\epsilon_{\textrm{\rm SA}} =\frac{P_{\rm out}}{2\pi\tau_{\rm sec}}= \bigg(\frac{a}{a_{\rm out}(1 - e_{\rm out}^2)}\bigg)^{3/2}\bigg(\frac{m_{\rm out}^2}{m_{\rm tot}m_{\rm bin}}\bigg)^{1/2}.
\label{eq:epsSA}
\end{equation}

The additional effective “corrected double averaging” potential in terms of the vector elements is
 
\begin{equation} \label{phi-sa}
\Phi_{\rm CDA}(\boldsymbol{j},\boldsymbol{e}) = -\epsilon_{\rm SA}\Phi_0\bigg(\phi_{\rm circ}(j_z,e_z,e) + e_{\rm out}^2\phi_{\rm ecc}(\boldsymbol{j},\boldsymbol{e})\bigg),
\end{equation}
which has an axisymmetric term $\phi_{\rm circ} (j_z,e_z,e)$ when the outer orbit is circular, and non-axisymmetric term $\phi_{\rm ecc}$ which is important when the outer orbit is significantly eccentric. Explicit expressions for both of these terms are given by \cite{Luo2016}. We will discard the $\phi_{\rm ecc}$ term, as we are interested in axisymmetric orbits, thus the SA potential is

\begin{equation}
\Phi_{\rm SA}(j_z,e_z,e) = -\epsilon_{\rm SA}\Phi_0\frac{27}{64}j_z\bigg[\frac{1 - j_z^2}{3} + 8e^2 - 5e_z^2\bigg],
\end{equation}

The vector elements are expressed in terms of the Kepler elements in a reference frame where the $\hat{\boldsymbol{z}}$ direction is along the outer angular momentum. \cite{Luo2016} showed that these corrections are consistent with N-body integration results (see e.g. their Fig. 5).

\subsection{Relaxing the test particle limit}

In the test particle limit, the traditionally conserved quantity is the $z$-component of the angular momentum, $j_z$. In the general case, a modified conserved quantity can be found from the conservation of the angular momentum  $\textbf{L}_{\rm tot} = \textbf{L}_{\rm in} + \textbf{L}_{\rm out}$ \citep{haim2018, anderson2017, Liu2018}. Consider the magnitude

\begin{equation}\label{eq:angularmomentum}
L_{\rm tot}^2 = L_{\rm in}^2 + L_{\rm out}^2 + 2L_{\rm in}L_{\rm out}\cos i.
\end{equation}

$L_{\rm out}$ and $e_{\rm out}$ are constant to quadrupole order. The ratio of the circular inner angular momentum to the outer angular momentum is defined as 
\begin{align}
\eta \equiv  \frac{\mu_{\rm in}}{\mu_{\rm out}}\bigg[ \frac{m_{\rm bin}a}{m_{\rm tot}a_{\rm out}(1 - e_{\rm out}^2)}\bigg]^{1/2},
\label{eq:eta}
\end{align}
where $\mu_{\rm in} = m_0 m_1/m_{\rm bin}$  and $\mu_{\rm out} = m_{\rm bin} m_{\rm out}/m_{\rm tot}$ are the reduced masses of the inner and outer binaries, respectively. Rewriting ${L}_{\rm in} = j\eta {L}_{\rm out}$, we can rewrite Eq.~\ref{eq:angularmomentum} in terms of the approximately conserved quantity $K_2$:

\begin{equation}\label{eq:K2}
    K_2 \equiv \frac{L_{\rm tot}^2 - L_{\rm out}^2}{2\eta L_{\rm out}^2} = j\cos i + \frac{j^2\eta}{2}.
\end{equation}
We can alternatively use 
\begin{equation}\label{K1}
K_1 = K_2 - \frac{\eta}{2} =j\cos i - \frac{e^2\eta}{2}.
\end{equation}

In the limit $L_{\rm in} \ll L_{\rm out}$, $\eta \rightarrow 0$ and $K_1 = K_2$ reduce to the familiar conserved quantity $j_z$. We take the initial eccentricity $e_0 \sim 0$, so that $K_1 \sim \cos i_0$.

We note that the procedure of \cite{Luo2016} corrects for the additional secular terms that arise due to $\epsilon_{\rm SA}$. However, there are also osculating orbital elements that vary on outer orbital timescales that do not affect the secular evolution. In other words, the quantity $K_2$ is conserved on average, but fluctuates around this mean value on a dynamical timescale.  Small  fluctuations do not affect the (corrected) secular evolution, but become important when the fluctuation $\delta e$ is larger than its mean value $e_{\rm max}$ (cf.~section \ref{sec-fluc}). 

\subsection{General Relativistic corrections}
When the separation between two bodies is sufficiently small, so that the gravitational radius $r_g = Gm_{\rm bin}/c^2$ is not negligible relative to $a (1-e)$, general relativity (GR) affects the binary's evolution.  Here we include the contribution of GR corrections to the interaction energy at the leading post-Newtonian (PN) order \citep{Blaes2002, Liu2015, Liu2018}:

\begin{equation}
    \Phi_{\textrm{GR}} = -\epsilon_{\textrm{GR}}\Phi_0\frac{1}{(1 - e^2)^{1/2}},\label{phi-gr}
\end{equation}

where

\begin{equation} 
\epsilon_{\textrm{GR}} = \frac{3m_{\rm bin}(1 - e_{\rm out}^2)^{3/2}}{m_{\rm out}}\bigg(\frac{a_{\rm out}}{a}\bigg)^3\frac{r_g}{a}, \label{eps_gr}
\end{equation}

is the ratio of GR to ZLK precession rates. In the limit $\epsilon_{\textrm{GR}} \gg 1$, GR precession is significant enough for the LK mechanism to be suppressed. For large precession $e_{\rm max} \rightarrow 0$, and the inner binary essentially evolves in isolation. If the inner separation is small enough, the binary may merge due to gravitational-wave (GW) dissipation within a Hubble time as an isolated binary \citep{Peters1964, Fragione2019}.

We note that similar estimates can be made for the impact of tidal interactions and rotational bulges \citep{Liu2015}.

\section{Computing the maximum eccentricity}\label{formalism}

The total interaction term in the specific energy is obtained by adding the contributions of Eqs.~ \ref{phi-quad}, \ref{phi-sa}, and \ref{phi-gr}: 

\begin{equation}\label{phi_tot}
    \Phi_{\rm tot} = \Phi_{\rm Quad}+\Phi_{\rm SA}+\Phi_{\rm GR}.
\end{equation}

The phase portrait under the Hamiltonian flow governed by the potential in Eq. \ref{phi_tot} is still one dimensional, and hence the dynamics are still integrable for a circular outer orbit, $e_{\rm out}=0$.

Similarly to \cite{Grishin2018}, we evaluate $\Phi_{\rm tot}$ at two points on the orbit, but we use the conservation of $K_2$ (instead of $j_z$) to account for the general case of non-negligible inner angular momentum ($\eta >0$), similarly to the approach taken by \citet{anderson2017} for the purely DA case ($\epsilon_{\rm SA}=0$). This unified approach leads to an implicit equation for $e_{\rm max}$, as shown below.

\begin{figure*}[htp!]
\centering
\includegraphics[width = \textwidth]{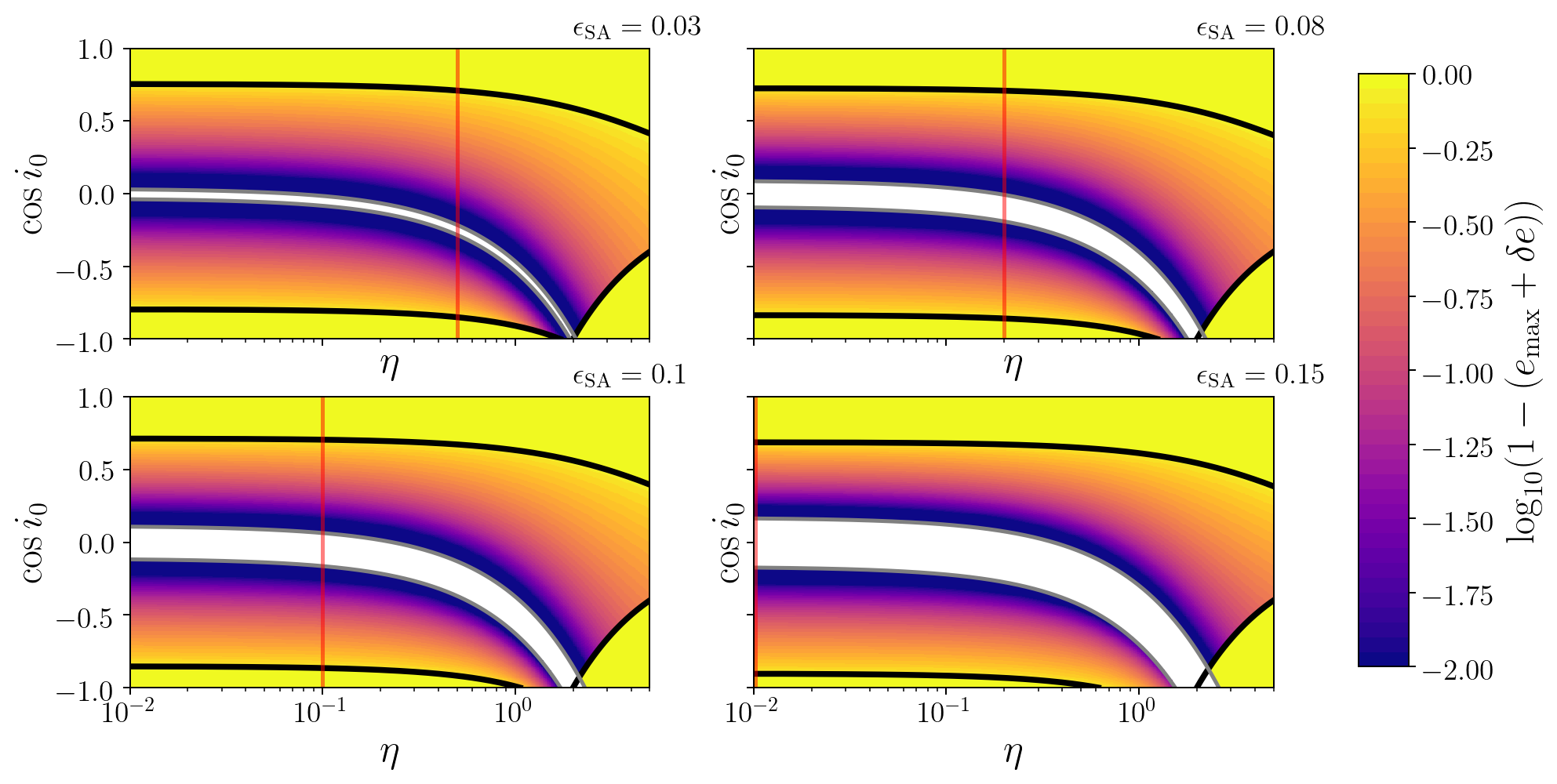}
\caption{The maximum eccentricity $e_{\rm max} + \delta e$ of the inner binary, initially inclined by angle $i_0$ relative to the outer binary, of a triple with inner-to-outer angular momentum ratio $\eta$ (see Eq.~\ref{eq:eta}), for four choices of the hierarchy parameter $\epsilon_{\rm SA}$ labeled in the top right corner.  The quantity $1 - (e_{\rm max}+\delta e)$ is plotted as a colour map with a logarithmic scale. The solid black lines indicate the boundaries of zero and non-zero eccentricity and are the same as in \citet{anderson2017} (see text). Purple colours indicate highly eccentric systems. The vertical red lines indicate fixed values of $\eta$ for which we carry out N-body numerical simulations. Regions with $e_{\rm max}+\delta e\ge1$ are shown in white.  The grey lines bound the region where the quasi-secular corrections break down, cf.~Eq.~(\ref{eq:non-secular-limit}).}
\label{fig:Newtoniancontour}
\end{figure*}

For the initial point on the phase portrait, we specify $i_0$ and $e_0$, and set $\omega = 0$. The latter choice of $\omega = 0$ places the initial condition in a circulating orbit as long as $e_0>0$\footnote{$e_0=0$ and $\omega=0$ is an unstable (hyperbolic) fixed point that lies on the separatrix.}. The potential at this point is

\begin{align}
\frac{\Phi_1}{\Phi_0} = &\frac{1}{8}\bigg[1 - 6e_0^2 - 3(1 - e_0^2)\cos^2i_0\bigg] - \epsilon_{\rm GR}\frac{1}{j_0} \nonumber \\
 -&\epsilon_{\rm SA} \frac{27}{64}j_z\bigg(\frac{1 - j_z^2}{3} + 3e_0^2 + 5e_0^2\cos^2 i_0\bigg), 
\end{align}
where $j_0=\sqrt{1-e_0^2}$, and $j_z = j_0 \cos i_0$ is evaluated at its initial value. 

The second point is evaluated at $e_{\rm max}$, which is attained at $\omega = \pm \pi/2$ for axisymmetric potentials \citep[e.g.][]{hamilton2}.  The potential at this point is  

\begin{align}
\frac{\Phi_2}{\Phi_0} = &\frac{1}{8}[1 + 9e^2_{\rm max} - 3j_{\rm min}^2\cos^2 i_{\rm m} - 15e^2_{\rm max}\cos^2i_{\rm m}] \nonumber \\ - &\frac{27\epsilon_{\rm SA}}{64}j_{\rm min}\cos i_{\rm m}\bigg( \frac{1 - j^2_{\rm min}\cos^2 i_{\rm m}}{3} + 3e^2_{\rm max}\nonumber  \\ + &5e^2_{\rm max}\cos^2 i_{\rm m}\bigg) - \epsilon_{\rm GR}\frac{1}{j_{\rm min}}, 
\end{align}

where $j_{\rm min} =\sqrt{1-e_{\rm max}^2}$ and $\cos i_{\rm m}$ is the mutual inclination at $e=e_{\rm max}$. We assume a small initial eccentricity ($e_0 \ll 1$) in order to focus on how circular binaries become eccentric.  Conservation of $K_1$, Eq.~(\ref{K1}), yields $\cos i_{\rm m} = (K_1 + e^2_{\rm max}\eta/2)/j_{\rm min}$, where $K_1 \approx \cos i_0$. Equating $\Phi_1=\Phi_2$ allows us to find an implicit expression for $e_{\rm max}$ in the appropriate ZLK window \citep{anderson2017}:
\begin{align}
    0=&-5g_{\rm max}^2 + j_{\rm min}^2(3 - \cos i_0\eta -\frac{\eta^2}{4}e_{\rm max}^2) \nonumber \\
    &+ \frac{3\epsilon_{\rm SA}}{8}\bigg[ -\frac{j_{\rm min}^2\cos^3i_0}{e_{\rm max}^2} - \frac{\eta j_{\rm min}^2}{2} -9j_{\rm min}^2 g_{\rm max} \nonumber \\ &+g_{\rm max}^3\bigg(\frac{1}{e_{\rm max}^2} - 16\bigg)  \bigg] +  \frac{8\epsilon_{\rm GR}j_{\rm min}}{3e_{\rm max}^2}\left( j_{\rm min} - 1\right),\label{eq:emax}
\end{align}
where $g_{\rm max}=\cos i_0 + \eta e_{\rm max}^2/2$. Eq.~(\ref{eq:emax}) allows us to extract (albeit implicitly) $e_{\rm max}$ in terms of $\epsilon_{\rm SA}, \eta, i_0$ and $\epsilon_{\rm GR}$, which are all fixed and determined from initial conditions.

In the test particle limit, without corrected DA and in the purely Newtonian regime, the usual expression \citep{Naoz2016} for the maximal eccentricity is obtained:

\begin{equation}\label{maxDA}
e^{\rm DA}_{\rm max} = \sqrt{1 - \frac{5}{3}\cos^2 i_0}.
\end{equation}

\subsection{Fluctuation term} \label{sec-fluc}

We have derived an expression for $e_{\rm max}$ by equating the potential, including the $\epsilon_{\rm SA}$ term, at the initial and maximum-eccentricity points, assuming a conserved $K_2$. However, $K_2$ is only conserved on average, and actually fluctuates around this averaged value, where the fluctuating amplitude is \citep{haim2018}

\begin{equation}\label{eq:dK21}
|\Delta K_2| = (1 + |K_2|\eta) \Delta j_z.
\end{equation}
This results in a fluctuating eccentricity ($\delta e$) about the mean value ($e_{\rm max}$). The total maximal eccentricity is
\begin{equation}\label{eq:ecorr}
e_{\rm max} + \delta e = \sqrt{1 - (j_{\rm min} - \delta j)^2},
\end{equation}
where $\delta j$ is the fluctuation in $j$. 

Differentiating $K_2$ from Eq.~(\ref{eq:K2}), we find
\begin{equation}\label{eq:dK22}
\Delta K_2 = (\cos i_{\rm m} + j \eta)\delta j.
\end{equation}
The inclination $i_{\rm m}$ at which $e_{\rm max}$ is attained falls on the solid black lines in Fig. \ref{fig:Newtoniancontour}. We focus on configurations where $\eta \lesssim 0.5$ (see Sec.~\ref{astro}), and we approximate $\cos i_m$ as $\pm \sqrt{3/5}$. The next contribution is of order $\sim \eta/10 \ll \mathcal{O}(1)$, so it can be neglected.

The maximal fluctuation in $j_z$ is \citep{haim2018,Grishin2018} 

\begin{equation}
\Delta j_z = \frac{15\epsilon_{\rm SA}}{8}  e^2_{\rm max} \cos^2 i_{\rm m}. \label{djz}
\end{equation}

Using  Eqs.~(\ref{eq:dK21}), (\ref{eq:dK22}) and (\ref{djz}) at $e_{\rm max}$, $\delta j$ is 

\begin{align}
\delta j = \frac{1 +  |K_2|\eta}{\sqrt{3/5} + j_{\rm min} \eta} \frac{9}{8}\epsilon_{\rm SA} e_{\rm max}^2\equiv  \mathcal{C} \epsilon_{\rm SA} e_{\rm max}^2.\label{delta_j}
\end{align}

Finally, for the fluctuating eccentricity, we square Eq.~(\ref{eq:ecorr}), neglect the $(\delta e)^2$ term (assuming that $e_{\rm max} \gg \delta e$), and substitute our expression for $\delta j$ from Eq. (\ref{delta_j}):

\begin{align}\label{eq:fluctuation}
\delta e = &\frac{j_{\rm min}}{e_{\rm max}}\delta j - \frac{(\delta j)^2}{2e_{\rm max}} \nonumber \\
		= &\mathcal{C}\epsilon_{\rm SA} e_{\rm max}\left( j_{\rm min} - \frac{e_{\rm max}^2 }{2} \mathcal{C}\epsilon_{\rm SA} \right).
\end{align}

The fluctuating eccentricity is a function of $i_0,\eta,\epsilon_{\rm SA}$, and $e_{\rm max}$, which are all determined from initial conditions. For a specific domain of initial inclination the fluctuation becomes sufficiently large such that the maximal eccentricity $e_{\rm max} + \delta e \rightarrow 1$ and the binary merges or becomes unstable.

The treatment leading to Eq.~(\ref{eq:fluctuation}) is valid for mildly hierarchical triples with small oscillations $\delta e \ll e_{\rm max}$.  The regime of validity of this treatment can be determined by considering considering when $|K_2| \leq |\Delta K_2|$, at which time the perturbations become so large that the eccentricity is not well defined: 
\begin{equation}
\left| \cos i_0 + \frac{\eta}{2} \right| \leq \frac{\frac{9}{8}\epsilon_{\rm SA}e_{\rm{max}}^2}{1 - \frac{9}{8}\epsilon_{\rm SA}e_{\rm{max}}^2\eta}
\approx \frac{\frac{9}{8}\epsilon_{\rm SA}}{1 - \frac{9}{8}\epsilon_{\rm SA}\eta}.
\label{eq:non-secular-limit}
\end{equation}

\begin{figure}[htp!]
\centering
\includegraphics[width = 0.5\textwidth]{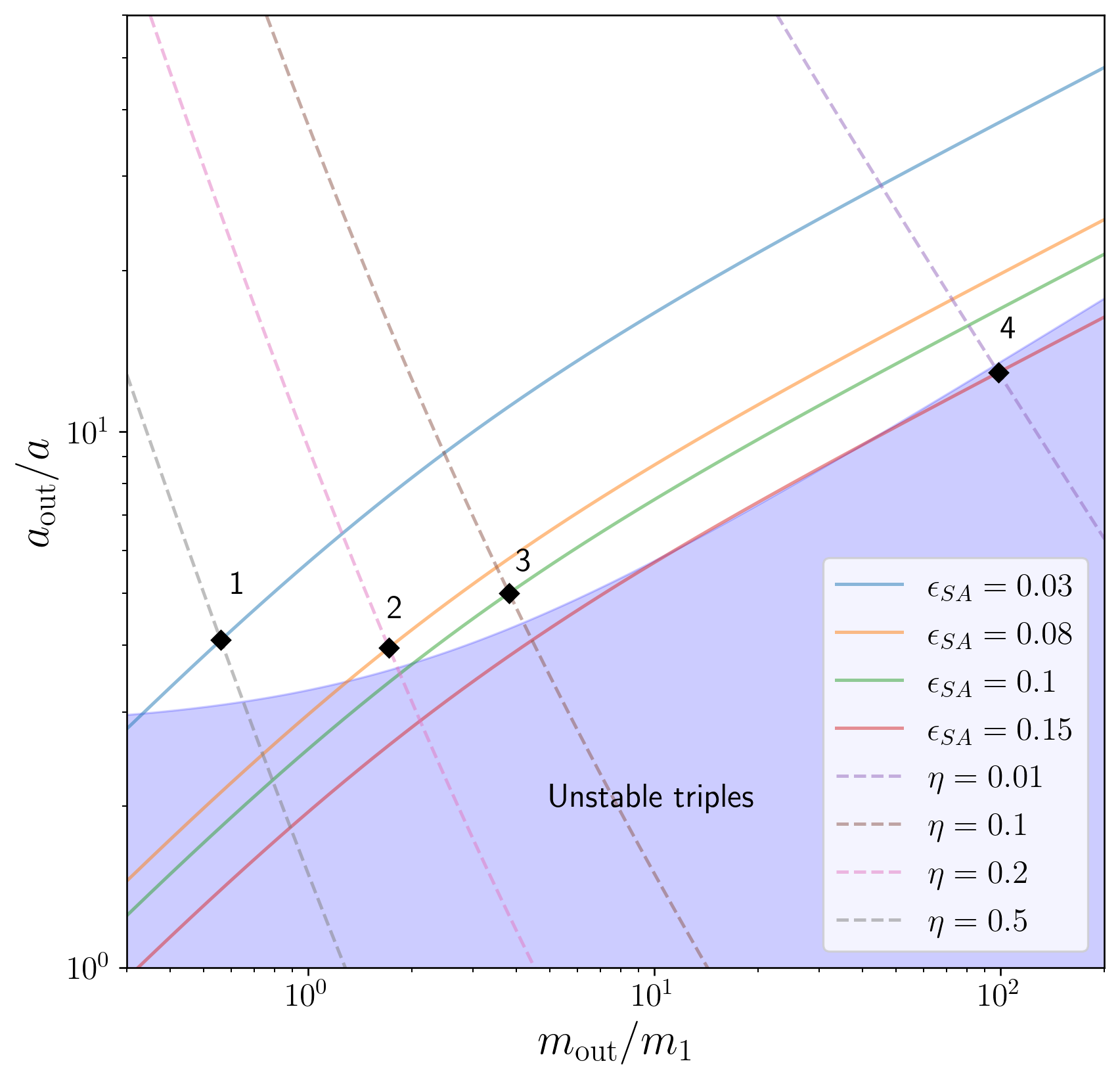}
\caption{Initial conditions in semi-major axis ratio and mass ratio space ($m_0 = m_1$). Dashed lines indicate fixed angular momentum ratio $\eta$ (Eq.~\ref{eq:eta}). Solid lines indicate fixed hierarchy parameter $\epsilon_{\rm SA}$ (Eq.~\ref{eq:epsSA}). Black diamonds denote initial conditions used for the set of N-body simulations, numbered as in Table~\ref{table:ICs}. Systems in the shaded lilac region are dynamically unstable according to the \citet{ma01} criterion.}
\label{fig:ICs}
\end{figure}

We note that in the test particle limit ($\eta \rightarrow 0$), the RHS of Eq.~\ref{eq:non-secular-limit} reduces to $(9/8)\epsilon_{\rm SA}$ as determined by \citet{Grishin2018}. 

\begin{figure*}[htp!]
\centering
\includegraphics[width = \textwidth]{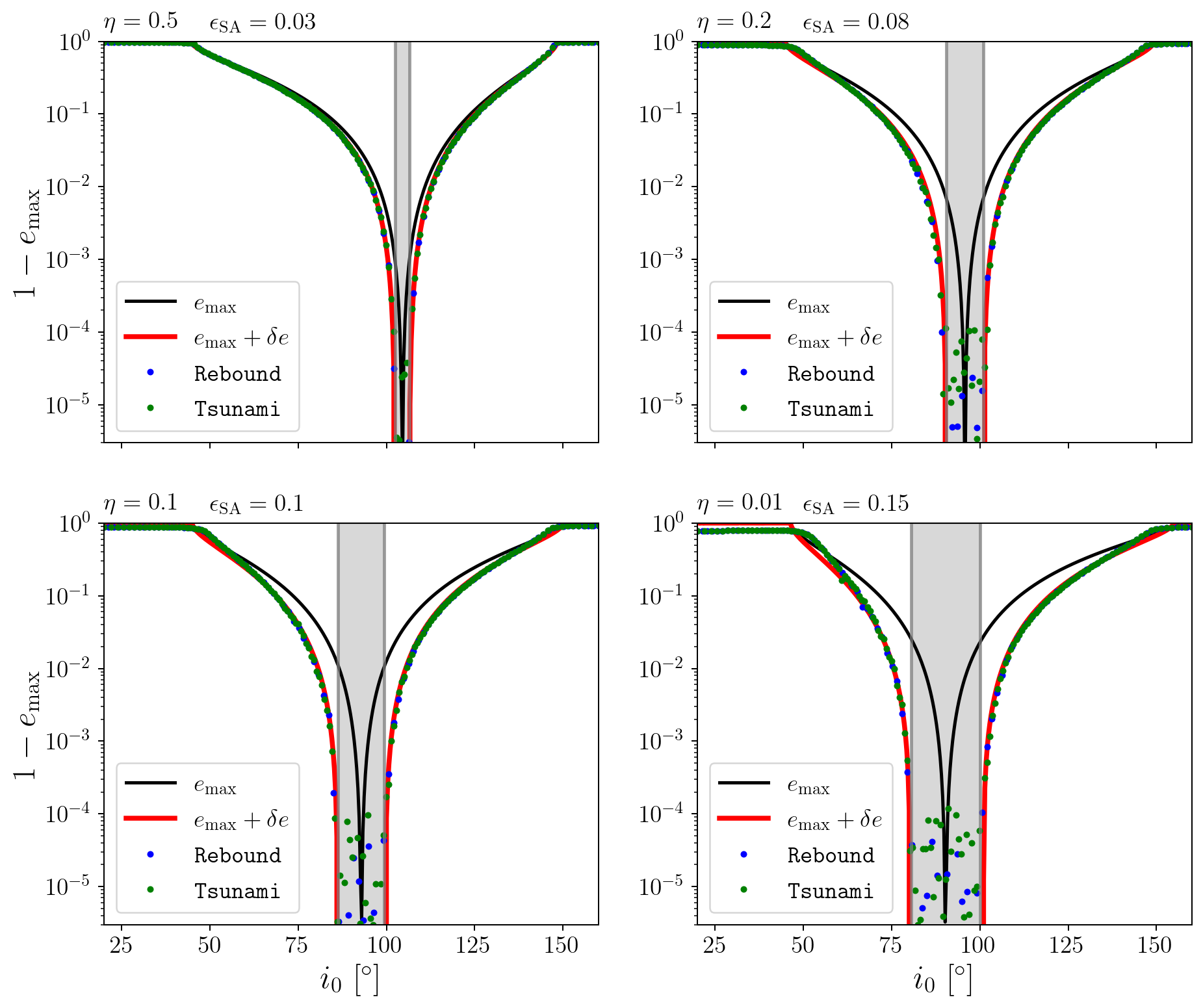}
\caption{Comparison of analytical $e_{\rm max}$ (black) and $e_{\rm max} + \delta e$ (red) prescriptions against N-body numerical simulations using \textsc{rebound} \citep[][blue dots]{rebound} and \textsc{tsunami} \citep[][green dots]{tsunami}. The grey region shows where $\Delta K_2 > K_2$, i.e., the eccentricity approaches unity, leading the inner binary to merge or the system to become unstable. The initial conditions are shown in Table~\ref{table:ICs}. The end time of all runs is 500 times the outer orbital period. In the top left panel we have also included the analytical $e^{\rm DA}_{\rm max}$ (black dashed) (Eq.~\ref{maxDA}), to show how DA neglects to account for large $\eta$ and $\epsilon_{\rm SA}$.}
\label{fig:Newtonian3}
\end{figure*}

We have developed an expression for $e_{\rm max}$ (Eq.~\ref{eq:emax}) and the fluctuation magnitude $\delta e$ (Eq.~\ref{eq:fluctuation}) relative to $e_{\rm max}$. Fig.~\ref{fig:Newtoniancontour} shows the maximum eccentricity $e_{\rm max} + \delta e$ of the inner binary, with initial mutual inclination $i_0$ to the tertiary body with inner-to-outer angular momentum ratio $\eta$ (Eq.~\ref{eq:eta}). The quantity $1 - (e_{\rm max} + \delta e)$ is plotted as a colour map on a logarithmic scale, for various choices of $\epsilon_{\rm SA} = \{0.03, 0.08, 0.1, 0.15\}$. The vertical red lines indicate specific choices of $\eta$ to compare against N-body simulations (see Fig.  \ref{fig:ICs}, \ref{fig:Newtonian3}). The black lines indicate the boundaries of zero and non-zero eccentricity. The solid lines are the same as in Fig. 1 of \citet{anderson2017}. 
The region between solid grey lines shows where the fluctuation becomes large and the eccentricity will approach unity ($\Delta K_2 > K_2$). In the limit of $\epsilon_{\rm SA}\to 0 $ we expect to reproduce the top left panel of \cite{anderson2017}, corresponding to $\epsilon_{\rm GR}=0$ (see also next section).
In the limit of $\eta \to 0$ we get back the results of \cite{Grishin2018}.

We are now ready to test our Newtonian analytic theory and relativistic extension via numerical simulations in the next section.

\begin{table}[]
\centering
\begin{tabular}{|l|l|l|l|l|l|}
\hline
  & $a$ [AU]    & $m_{\rm out}$ [$M_{\odot}$]    & $a_{\rm out}$ [AU]    & $\epsilon_{\rm SA}$ & $\eta$  \\ \hline
1 & 7.256 & 27.97 & 29.63 & 0.03    & 0.5  \\ \hline
2 & 2.151 & 85.61 & 8.510 & 0.08    & 0.2  \\ \hline
3 & 1.953 & 190.5 & 9.765 & 0.10    & 0.1  \\ \hline
4 & 1.295 & 4939. & 16.72 & 0.15    & 0.01 \\ \hline
\end{tabular}
\caption{Initial conditions corresponding to the chosen points in $(\epsilon_{\rm SA},\eta)$ space in Fig.~\ref{fig:ICs}.  We set $m_0 = m_1 = 50 M_{\odot}$ and $\epsilon_{\rm GR}=0.0001$, with the inner semi-major axis $a$, tertiary mass $m_{\rm out}$ and semi-major axis $a_{\rm out}$ following from the chosen $(\epsilon_{\rm SA},\eta)$.}
\label{table:ICs}
\end{table}

\section{Numerical methods and results}\label{results}

\subsection{Numerical codes}

We now seek to compare the analytic formalism we have developed in previous sections to N-body integration. We employ two different numerical integrators: \textsc{rebound} \citep{rebound} and \textsc{tsunami} \citep{tsunami}. Within \textsc{rebound}, we use \textsc{ias15}, a fast, adaptive, higher-order integrator for gravitational dynamics, accurate to machine precision over a billion orbits \citep{reboundias15}. \textsc{tsunami} is a fast and accurate few-body code designed to follow the evolution of self-gravitating systems. The integrator is based on \cite{Mikkola1999} algorithmic regularisation and can easily handle close encounters, highly hierarchical systems, and extreme mass ratios. \textsc{tsunami} also includes post-Newtonian (PN) corrections through 2.5 PN order, i.e., including both relativistic precession and lowest-order radiation reaction, allowing us to explore systems which have large $\epsilon_{\rm GR}$.  We set a collision radius equal to five times the Schwarzschild radius of the particles in order to avoid unphysical behavior when the PN expansion breaks down.  Whenever the distance between two particles is less than the sum of their collision radii, the simulation is stopped. Because the triples we consider are close to the instability regime (see \autoref{fig:ICs}), it is possible that some systems will break up over the course of the simulation. We stop a simulation and label it as unstable if the binding energy of either the inner or outer binary becomes negative. Otherwise, the end time of all runs is 500 times the outer orbital period.

\begin{figure*}[htp!]
\centering
\includegraphics[width = \textwidth]{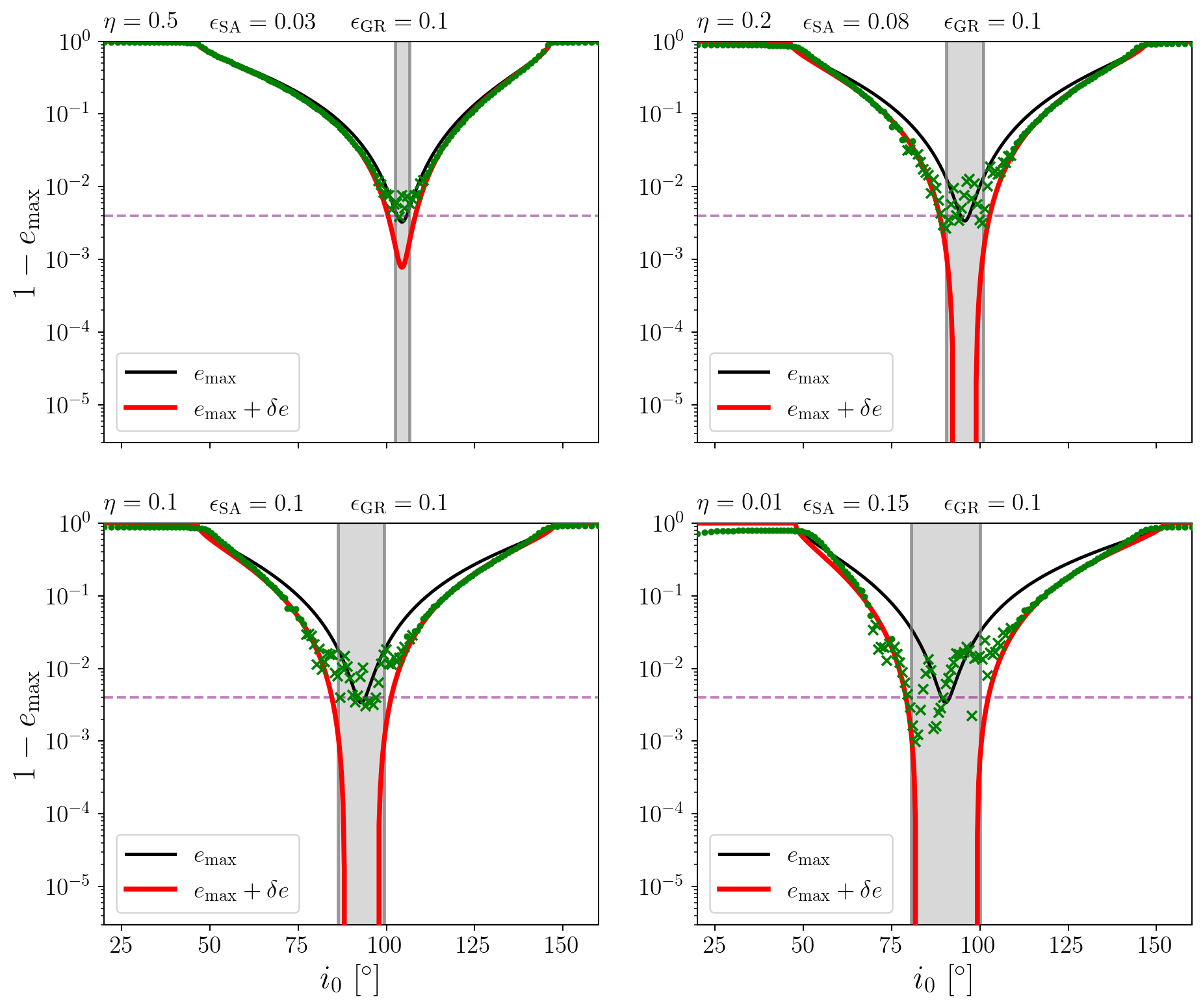}
\caption{Comparison of analytical $e_{\rm max}$ (black) and $e_{\rm max} + \delta e$ (red) prescriptions against N-body numerical simulations using \textsc{tsunami} \citep[][green dots, crosses]{tsunami}. Dots are indicative of initial inclinations for which the system remains stable. Crosses indicate initial inclinations that lead to collision or break-up of the inner binary.  The grey region shows where $\Delta K_2 > K_2$, i.e., the eccentricity is formally unconstrained, leading the inner binary to merge or the system to become unstable. The initial conditions follow Table~\ref{table:ICs}, with both separations scaled by a factor of $10^{-3}$ such that $\epsilon_{\rm GR} = 0.1$ in all cases. The purple dashed line is the expected limit at small $\epsilon_{\rm GR}\ll1$, $e_{\rm max} \approx \sqrt{1-(8\epsilon_{\rm GR}/9)^2}\approx 0.996$.}
\label{fig:GR3}
\end{figure*}

\subsection{Initial conditions}

The set of initial conditions we have chosen to compare against N-body integrations are shown in Fig.~\ref{fig:ICs} and Table~\ref{table:ICs}. Fig.~\ref{fig:ICs} shows initial conditions in semi-major axis ratio and mass ratio space for $m_0 = m_1$. Dashed lines indicate fixed angular momentum ratio $\eta$ (Eq.~\ref{eq:eta}). Solid lines indicate fixed hierarchy parameter $\epsilon_{\rm SA}$ (Eq.~\ref{eq:epsSA}). Black diamonds denote the set of N-body simulations, numbered as in Table~\ref{table:ICs}. The shaded lilac area indicates dynamically unstable systems following the criterion of \citet{ma01}:

\begin{equation}
\frac{a_{\rm out}}{a} \leq 2.8\bigg[ \bigg( 1 + \frac{m_{\rm out}}{m_{\rm bin}} \bigg) \bigg]^{2/5}.
\label{eq:mardling}
\end{equation}

For non-negligible values of both $\epsilon_{\rm SA}$ and $\eta$, the systems must be compact and close to the stability boundary. The \citet{ma01} stability limit is too strict for $m_{\rm out}\gg m_1$, so simulation 4 is still stable.

For a concrete example, we consider an equal mass inner binary of $50M_\odot$ BHs. The separation of the inner and outer binaries and mass of the tertiary body are then determined from the choice of $\eta$, $\epsilon_{\rm SA}$ and $\epsilon_{\rm GR}$. All simulations start with $e = 0$, $ \Omega = \pi/4$, $ \omega = \pi /2$ and $f = 0$. The outer orbit has $e_{\rm out} = 10^{-5}$, $ \Omega_{\rm out} = \omega_{\rm out} = f_{\rm out} = 0$. $\Omega$ is the longitude of ascending node. $f$ is the true anomaly. This ensures that the osculating $j_z$ is at its mean value.

\subsection{Newtonian case}

Fig. \ref{fig:Newtonian3} shows the comparison of our analytic prescription against numerical integrations. The initial conditions are the red vertical slices in the corresponding panels of  Fig.~\ref{fig:Newtoniancontour}. The analytic curves $e_{\rm max}$ (black) and $e_{\rm max} + \delta e$ (red) are compared against N-body integrations computed using \textsc{rebound}  (blue) and \textsc{tsunami}  (green). The simulations closely follow the $e_{\rm max} + \delta e$ curve for a range of values of $\epsilon_{\rm SA}$ and $\eta$. Furthermore, N-body simulations match the region we have determined (Eq.~\ref{eq:non-secular-limit}) within which the eccentricity becomes unconstrained (grey region). The width of the grey region is proportional to the magnitude of $\epsilon_{\rm SA}$, and the grey region appears to be positioned at larger initial inclinations for larger values of $\eta$. 

\subsection{Including GR effects}\label{greffects}

We now consider systems with $\epsilon_{\rm GR} = 0.1$, shown in Fig.~\ref{fig:GR3}. In order to achieve such a large value of $\epsilon_{\rm GR}$, the inner separation is of order $10^{-3} \rm AU$, so even a circular binary with $50 M_\odot$ components will merge in $\sim 1$ year through GW emission \citep{Peters1964}.  On the other hand, the merger may be eccentric, hence it is instructive to study this case.

Comparing Figs.~\ref{fig:Newtonian3} and \ref{fig:GR3}, we notice that $e_{\rm max}$ curves (black) only reach values of order $\sqrt{1- (8 \epsilon_{\rm GR}/9)^2} \approx 0.996$ in the GR case as relativistic precession suppresses the ZLK mechanism, compared to eccentricity approaching unity in the Newtonian case (cf.~Eq.~(\ref{eq:emax}) with $\epsilon_{\rm SA}=\eta=0$). This is consistent with the result of \cite{Liu2015}.

We notice similar behaviour when examining the $e_{\rm max} + \delta e$ (red) curve, leading to a ``shrinking'' of the region where $\Delta K_2 > K_2$. The criterion we have developed (Eq.~\ref{eq:non-secular-limit}) no longer follows the regions within which the eccentricity becomes unconstrained. In the case of $\eta = 0.5, \epsilon_{\rm SA} = 0.03$, the red curve indicates that the grey region ought to disappear entirely as GR effects prevent the inner binary from flipping.

Some orbits, especially the highly eccentric ones, merge through GW emission or come sufficiently close to be labeled as collisions (green crosses in Fig. \ref{fig:GR3}) within the duration of the simulation. At an eccentricity of 0.99, the merger timescale shrinks by a factor of almost one million relative to a circular binary, so that a binary consisting of two $50 M_\odot$ BHs at a separation of $0.001\ \rm AU$ merges in approximately 1 minute \citep{Peters1964, Mandel:2021}.  Binaries that reach more moderate maximal eccentricities will be circularised prior to the merger.

Overall, we find good correspondence between the analytic and numerical results. The only discrepancy is that occasionally maximal eccentricities appear to be slightly larger than predicted analytically. We find no difference in maximal eccentricity between 1PN only and full 2.5PN expansions, although energy dissipation through GW emission in the 2.5 PN expansion brings the system to merger.

\section{Discussion}\label{discussion}

\subsection{Astrophysical implications} \label{astro}

{\bf General picture:}
We derive and validate a new formula for the maximal eccentricity of the inner binary, implicitly given in Eq.~(\ref{eq:emax}). 

Perhaps the main effect is the expansion of the parameter space in which mildly hierarchical systems can reach very high eccentricity.  \citet{Grishin2018} discussed this in the test particle limit, and we find the same effect for triples of comparable masses.  In the hierarchical, test-particle, Newtonian limit, when $\epsilon_{\rm SA}=\epsilon_{\rm GR}=\eta=0$, Eq.~(\ref{eq:emax}) yields a range $|\cos i_0|<\sqrt{(3/5) (1-e_{\rm max}^2)}$ of initial inclinations that can yield eccentricities $e_{\rm max}$ or greater.  This range in $\cos i_0$ is only $\approx 0.07$ for $e_{\rm max}>0.999$, as illustrated by the $e^{\rm DA}_{\rm max}$ curve in the top left panel of Fig. \ref{fig:Newtonian3}.  On the other hand, relaxing the hierarchical limit, Eq.~(\ref{eq:non-secular-limit}) predicts that arbitrarily large eccentricities can be reached for $|\cos i_0|< (9/8) \epsilon_{\rm SA}$ (with $\epsilon_{\rm GR}=\eta=0$) -- i.e., a range of $0.34$ in $\cos i_0$ for $\epsilon_{\rm SA}=0.15$.  This is illustrated by the thickness of the grey region in the bottom left panel of Fig. \ref{fig:Newtonian3}.  Thus, high maximal eccentricities are significantly more likely for mildly hierarchical systems, assuming an isotropic (flat in $\cos i_0$) distribution of initial inclination angles between the inner and outer orbits.

The extension beyond the test particle limit shifts the initial inclinations that lead to $e_{\rm max} \to 1$ toward retrograde configurations. The range of cosines of initial inclination angles leading to unbound eccentricity is increased by a fraction $(1-9\epsilon_{\rm SA}\eta/8)^{-1}$ (see Eq. \ref{eq:non-secular-limit}), but it is a mild correction in most cases. 

The extreme case of $\eta \gg 1$ requires the tertiary to be much less massive than the inner binary, thus the inner binary is essentially unperturbed. This is the "inverse ZLK problem" \citep{naoz2017} and is not discussed here.

We tested our results both with Newtonian and PN codes, where the strength of GR is encapsulated in $\epsilon_{\rm GR}$ (Eq.~\ref{eps_gr}). It is important to stress that this is the {\it relative} strength of GR precession to ZLK precession, and not a statement on the proximity to the gravitational radius and the breakdown of the PN expansion. 

If all three masses are similar and the outer orbit is circular, Eq.~(\ref{eq:epsSA}) reduces to $\epsilon_{\rm SA} \sim (a/a_{\rm out})^{3/2}$, Eq.~(\ref{eq:eta}) to $\eta \sim (a/a_{\rm out})^{1/2}$, and Eq.~(\ref{eps_gr}) to $\epsilon_{\rm GR} \sim (a/a_{\rm out})^{-3} (r_g/a) $.  Stability requires that $a_{\rm out}$ is at least a few times larger than $a$, so in this regime, non-negligible $\eta$ and $\epsilon_{\rm SA}$ are generally obtained when $a$ is of order $0.1 a_{\rm out}$.  A significant $\epsilon_{\rm GR}$ then requires $a \lesssim 10^4 r_g$.  Once eccentricity is driven up to large values by ZLK resonances, the inner binary periapsis in such systems becomes sufficiently small that they merge rapidly through GW emission.  This can leave very limited opportunities for interactions that put systems into this regime of interest.

{\bf Stellar BH and stars:} For the case of stellar mass BHs, in order for for $\epsilon_{\rm SA}, \eta$ and $\epsilon_{\rm GR}$ to all be non-negligible, requires extremely compact systems that merge within thousands of years.  Isolated field triples could fit into a compact system for $a_{\rm out} \gtrsim 100 R_\odot$ \citep{vg_t21}, which corresponds to  $\epsilon_{\rm GR} \sim \mathrm{few} \times 10^{-3}$, so the treatment can still be Newtonian. However, such compact triples are expected to be coplanar due to interactions during stellar evolution. Stellar evolution may weaken the hierarchy of otherwise stable triples \citep{PeretsKratter:2012,toonen21}. Such triples will enter the semi-secular regime before becoming dynamically unstable, which may result in an enhanced rate of mergers and collisions. Dense environments such as globular clusters can in principle construct such compact triples and lead to {\it eccentric} mergers. \citep{Fragione2019}, as potentially inferred for GW190521 \citep{isobel2020}.

{\bf SMBHs:} Supermassive triple BHs are more promising in having larger values of $\epsilon_{\rm GR}$ for longer, since the GW-driven inspiral duration scales with the mass when the separation in units of the gravitational radius, $a/r_g$, is fixed.  This leaves more time for dynamical or gas interactions to insert systems into the regime of interest.  \cite{bonetti16} show that $\sim 10^8 M_\odot$ BHs may stall at radii of around a pc and triple-induced mergers are plausible. The fiducial system in their Fig.~6 has $\epsilon_{\rm GR} \approx 0.7,\ \eta \approx 0.2,\ \epsilon_{\rm SA} \approx 0.02$. In a follow-up Monte Carlo study,  \cite{bonetti2} found that between $20-30\%$ of the initially sampled triple BH systems are merging, with preference for equal mass inner binary (see their Fig.~6.). Without PN evolution, the merger rate is around $40-60\%$, roughly twice than the rate with PN terms. 

\subsection{Limitations and caveats}

{\bf Outer eccentricity:} We limited our study to outer circular binaries, $e_{\rm out}=0$. For non-zero eccentricity, the corrected averaging term has a non-axisymmetric contribution, hence $K_2$ is no longer conserved and the dynamics are chaotic. Moreover, for unequal inner masses, the octupole term may also induce chaotic evolution at longer timescales. In the DA test-particle limit, $e_{\rm max}$ is then unconstrained for a wider range of initial inclinations, depending on the octupole strength \citep[e.g][]{naoz11,katz2011,munoz2016}. We expect the maximal eccentricity to be also increasing compared to outer circular rbits on average. A systematic study is deferred for future work.

{\bf Tides, stellar evolution and other complications:} 
A similar treatment can account for tides when the inner binary contains a non-degenerate star. Equilibrium tides \citep{hut81} cause extra precession and can also quench ZLK oscillations, similarly to GR. The strength of these tides depends on the radii and apsidal motion constants of the stars, and $e_{\rm max}$ had been calculated analytically in the DA regime \citep{Liu2015}. It is possible to extend the study to main sequence or giant stars which have a convective envelope. Similarly to GW dissipation, tidal dissipation could also be important and reduce the separation. We do not consider it here. 

\subsection{Future work}
 The interplay of eccentricity, spin and orbital apdisal precession affects the last stages of compact BH evolution \citep{phukon19}. It is possible that the librating argument of pericentre around $\pi/2$ in the ZLK regime could serve as a physically motivated prior and improve eccentric GW templates for the analysis of mergers from dynamical formation channels. 
 
 Analytical results can play a key role in other large-scale numerical modelling, such as population synthesis of triple systems. Our results could serve as a prescription and save the computational cost of  numerically integrating orbital dynamics.

\section{Summary} \label{sum}

We have derived an analytical expression for the maximal eccentricity of hierarchical triple systems with an outer circular orbit. The expression is valid for {\it any} angular momentum ratio and {\it any} orbital period ratio (provided that the system is dynamically stable), and can thus be applied to {\it any} triple system which satisfies the assumptions of a  circular outer orbit and dynamical stability. The main results are Eqs.~(\ref{eq:emax}), (\ref{eq:fluctuation}) and Fig.~\ref{fig:Newtoniancontour}. This contributes to the analytic understanding of the long term stability and evolution of triples of field stars or BHs of any mass, and can be used as a prescription in population synthesis codes.

After verifying and recovering previous results, we focus on mild hierarchy ($\epsilon_{\rm SA} \gtrsim 0.05$, Eq.~\ref{eq:epsSA}) {\it and} comparable masses ($\eta \gtrsim 0.1$, Eq.~\ref{eq:eta}). These systems are inherently close to being dynamically unstable (Fig. \ref{fig:ICs}). Nevertheless, we are able to obtain accurate results with good agreement with numerical simulations (Fig. \ref{fig:Newtonian3}, \ref{fig:GR3}) in both Newtonian and PN limits. Similarly to the test particle case, mild hierarchy (moderately large $\epsilon_{\rm SA}$) is mainly responsible for the increase in $e_{\rm max}$ and the range of initial inclinations over which it can be achieved, while the angular momentum ratio $\eta$ mostly shifts the initial inclinations contributing to maximum eccentricities towards retrograde orbits.

We expect that GR corrections to triple dynamics are small for stellar mass BHs, and Newtonian treatments should be sufficient.  On the other hand, SMBH triples do require PN corrections. Other stars and compact objects may also experience enhanced rates of collisions or mergers via the ZLK mechanism in the mildly hierarchical, non-test-particle limit, though additional aspects of tides and stellar evolution need to be taken into account.

\section*{Acknowledgements}
We thank Isobel Romero-Shaw and Alejandro Vigna-G{\'o}mez for useful conversations. IM is a recipient of the Australian Research Council Future Fellowship FT190100574. AAT received support from JSPS KAKENHI Grant Numbers 19K03907 and 21K13914. This research was supported in part by the National Science Foundation under Grant No. NSF PHY-1748958. 

\bibliographystyle{aasjournal}
\bibliography{references} 
\end{document}